\documentclass[]{pasj01}
\draft
\usepackage{comment}
\usepackage{color}

\begin{document} 
\Received{}
\Accepted{}

\title{Universal detection of high-temperature emission in X-ray Isolated Neutron Stars}

\author{Tomokage \textsc{yoneyama}\altaffilmark{1,2}%
}
\altaffiltext{1}{Department of Earth and Space Science, Graduate School of Science, Osaka University, 1-1 Machikaneyama-cho, Toyonaka, Osaka 560-0043, Japan}
\altaffiltext{2}{Project Research Center for Fundamental Sciences, Graduate School of Science, Osaka University, 1-1 Machikaneyama-cho, Toyonaka, Osaka 560-0043, Japan}
\altaffiltext{3}{College of Science and Engineering, Kanto Gakuin University, 1-50-1 Mutsuurahigashi, Kanazawa-ku, Yokohama, Kanagawa 236-8501, Japan}

\email{yoneyama@ess.sci.osaka-u.ac.jp}

\author{Kiyoshi \textsc{hayashida}\altaffilmark{1,2}}

\author{Hiroshi \textsc{nakajima}\altaffilmark{3}}


\author{Hironori \textsc{matsumoto}\altaffilmark{1,2}}


\maketitle

\begin{abstract}
Strongly magnetized isolated neutron stars (NSs) are categorized into two families, according mainly to their magnetic field strength. The one with a higher magnetic field of $10^{14}$ - $10^{15}$ G is called magnetar, characterized with repeated short bursts, and the other is X-ray isolated neutron star (XINS) with $10^{13}$ G.
Both magnetars and XINSs show thermal emission in X-rays, but it has been considered that the thermal spectrum of magnetars is reproduced with a two-temperature blackbody (2BB), while that of XINSs shows only a single-temperature blackbody (1BB) and the temperature is lower than that of magnetars.
On the basis of the magnetic field and temperature, it is often speculated that XINSs may be old and cooled magnetars. 
Here we report that all the seven known XINSs show a high-energy component in addition to the 1BB model. Analyzing all the XMM-Newton data of the XINSs with the highest statistics ever achieved, we find that their X-ray spectra are all reproduced with a 2BB model, similar to magnetars.
Their emission radii and temperature ratios are also similar to those of magnetars except for two XINSs, which show significantly smaller radii than the others. The remarkable similarity in the X-ray spectra between XINSs and magnetars suggests that their origins of the emission are also the same. The lower temperature in XINSs can be explained if XINSs are older than magnetars. Therefore, these results are the observational indication that supports the standard hypothesis on the classification of highly-magnetized NSs.
\end{abstract}

\section{Introduction}
Two families of isolated neutron stars (NSs) that have unusually strong magnetic field and emit thermal radiation are observed in X-rays. The one is magnetars, and their magnetic field on the poles is estimated to be $B \sim 10^{14}$ –- $10^{15}$ G from X-ray pulsation driven by their rotation (Turolla, Zane and Watts 2015 for review). The other family is X-ray Isolated Neutron Stars (XINSs) with comparatively weak magnetic field of $B \sim 10^{13}$ G (Haberl 2007 for review).

Magnetars show persistent thermal X-ray spectra consisting of two-temperature blackbody (2BB) or blackbody with power-law at an energy band below 10 keV with a non-thermal hard tail above 10 keV (e.g. G\"{o}tz et al. 2006; Enoto et al. 2010, 2011). Temperatures of the 2BB spectra are typically 0.5 keV and 1 keV for cool and hot component, respectively with the X-ray luminosity $L_X$ of $10^{34}$ -- $10^{36}$ erg s$^{-1}$. Two types of NSs, Anomalous X-ray Pulsars (AXPs) and Soft Gamma-ray Repeaters (SGRs) are classified as magnetar. AXPs show significantly lower rotational energy loss than the thermal luminosity and thus called ``anomalous''. SGRs show short bursts repeatedly observed in hard X-ray/soft Gamma-ray band. The energy source of the bursts is considered to be release of magnetic energy, such as magnetic reconnection (Thompson and Duncan 1995; Corsi and Owen 2011). Some magnetars are associated with Super Nova Remnants (SNRs), which should be their born place (e.g. Kothes and Foster 2012). This is consistent with their young characteristic age ($10^{2}$ -- $10^{4}$ yr). Recently, some ``low field'' magnetars have been discovered. Their weak magnetic field of $B \sim 10^{13}$ G is consistent with that of XINS. However, they show characteristic bursts and thus categorized into magnetar.

XINSs have a comparatively weak magnetic field of $B \sim 10^{13}$ G. X-ray spectra of XINSs are generally  reproduced by a single-temperature blackbody model with a temperature of $kT \sim 100$ eV. Their X-ray luminosities are $10^{30}$ -- $10^{32}$ erg s$^{-1}$ (e.g., Tr\"{u}mper et al. 2004; Haberl 2007), and they also exceed the spin-down luminosity expected with $P \sim 3$ -- 11 s and $\dot{P} \sim 10^{-14}$ -- $10^{-13}$ s s$^{-1}$. Most XINSs have broad absorption lines in their spectra, and their origin is still under discussion (e.g. Borghese et al. 2017). So far, only seven XINSs discovered by ROSAT are known, so-called ``the Magnificent Seven (M7)''. The members are RX J0420.0$-$5022 (Haberl, Pietsch and Motch 1999; J0420 hereafter), RX J0720.4$-$3125 (Haberl et al. 1997; J0720), RX J0806.4$-$4123 (Haberl, Motch and Pietsch 1998; J0806), 1RXS J130848.6+212708 (Schwope et al. 1999; RBS1223), RX J1605.3+3249 (Motch et al. 1999; J1605), RX J1856.5$-$3754 (Walter, Wolk and Neuh\"{a}user 1996; J1856), and 1RXS J214303.7+065419 (Zampieri et al. 2001; RBS1774). Optical counterparts were identified only for the brightest two sources, J1856 (Walter and Matthews 1997) and J0720 (Kulkarni and van Kerkwijk 1998). Parallax measurement has been performed for the two sources to determine the distances. Walter et al. (2010) obtained $123^{+11}_{-15}$ pc for J1856 and Kaplan, van Kerkwijk and Anderson (2007) obtained $360^{+170}_{-90}$ pc for J0720. These are consistent with the distances estimated by interstellar X-ray absorption ($N_{\rm H} \sim 7 \times 10^{19}$ cm$^{-2}$ for J1856 and $1.2 \times 10^{20}$ cm$^{-2}$ for J0720; Posselt et al. 2007). They are considered to be cooling, steady source with characteristic ages of $\sim 10^6$ yr, apart from J0720, which exhibits long-term variation in timing and spectral properties (Hohle et al. 2012).

\begin{figure}[htbp]
 \begin{center}
   \includegraphics[width=150mm]{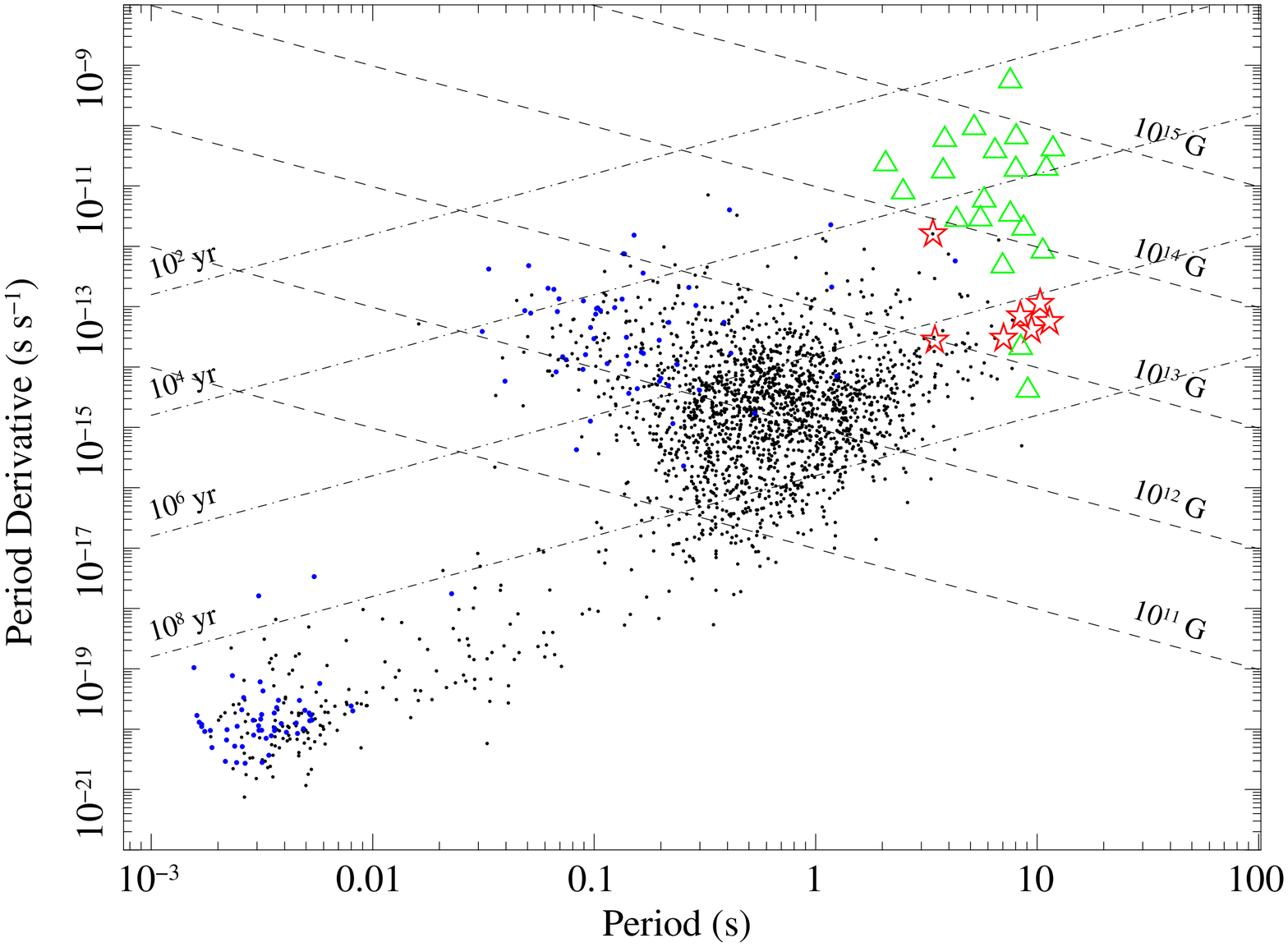}
    \caption{$P$ -- $\dot{P}$ relation of isolated NSs cited from the ATNF Pulsar Catalogue (Manchester et al. 2005, http://www.atnf.csiro.au/people/pulsar/psrcat/). The black dots indicate radio pulsars and the blue dots are radio pulsars with a high-energy emission. The red stars are XINSs and the green triangles are magnetars.}
    \label{fig:p-pdot}
  \end{center}
\end{figure}

In the period - period derivative ($P$ - $\dot{P}$) diagram, magnetars and XINSs distribute in adjacent locations; XINSs are $\sim 100 - 1000$ times older than magnetars and they have $1/10 - 1/100$ weaker magnetic field than magnetars (see figure \ref{fig:p-pdot}). In addition, as mentioned above, the surface temperature of XINS measured with X-ray observations are lower than that of magnetars. These facts lead to a hypothesis that XINSs are ``worn-out'' magnetars (e.g. Turolla 2009). Recently, theoretical study for magneto-thermal evolution of isolated NS also supports this hypothesis (Vigan\'{o} et al. 2013) . In this paper, we present another observational evidence to support this hypothesis but with suggesting a missing link between these two families of thermal radiating, strongly magnetize NSs..

All members of the M7 have been routinely observed over a decade since their discovery, particularly by the XMM-Newton satellite. We first studied J1856 (Yoneyama et al. 2017; hereafter referred to as Y17), the brightest and nearest ($\sim 120$ pc; Walter et al. 2010) one among them. It had been known that the X-ray spectrum of J1856 below 1 keV can be explained by two-temperature blackbody (2BB) component with temperatures of $kT \sim 63$ eV and 32 eV, the latter of which mainly accounts for the optical flux (Beuermann, Burwitz and Rauch 2006). In our study, we integrated all the spectra observed with Suzaku and XMM-Newton to achieve the maximum statistics with the available data. The integration is crucial to study the intrinsically faint high-energy end of the spectra, and we discovered an excess emission over the two-temperature blackbody model at the high-energy end, around 1 keV. We named this ``keV-excess''.  In this paper, prompted by our previous discovery, we search for and investigate, if found, the similar excess emission in the other six sources of the M7, using the XMM-Newton data. In section 2, we describe the data reduction, the examination for the long-term stability of the sources in order to make integrated spectra, and the X-ray spectral analysis using the single-temperature blackbody model. The analysis clarifies the existence of an excess emission similar to that of J1856 in the spectra of the other XINSs. In section 3, we quantify it and evaluate the systematic uncertainties. In section 4, we try to approximate the keV-excess emission for all of the M7, including J1856, and this analysis shows that the 2BB model can reproduce all X-ray spectra of the M7. In section 5, we compare the spectral parameters of the M7 with those of magnetars, showing the remarkable similarity between them which support the ``worn-out'' hypothesis, discuss the origin of the keV-excess, and speculate an evolution scenario of the strongly magnetized NSs.

Unless otherwise noted, the errors are $1\sigma$ in spectral plots and 90 \% confidence level in all other values. In this paper, we use ``gauss'' model in XSPEC to reproduce absorption line features. We confirmed our fitting results do not significantly altered if we used multiplicative model ``gabs'' model, usually used one.

\section{Data reduction and universal discovery of the keV-excess emission from the M7}

\subsection{Data reduction}
The details of the data reduction and examining systematic errors for J1856 are described
in Yoneyama et al. (2017, hereafter Y17). Here we describe those for the other six sources. For each object, we use the data observed by XMM-Newton EPIC-pn from 2000 to 2016, listed in table \ref{tab:pn_data}, together with J1856 adopted from Y17. We do not use the observations in which background flares occupied most of the exposure time. We do not use the EPIC-MOS data because of their long-term instability (Read et al. 2005). All of these observations were performed with the thin filter. We process raw event data with evselect of XMM-Newton science analysis package (SAS) v16.1.0, selecting only single and double events, excluding low pulse-height events and hot pixels. Then we perform Good Time Interval (GTI) selection to exclude background flares with \verb'espfilt' of SAS for the full-flame mode and \verb'XSELECT V2.4d' for the small-window mode, because espfilt is applicable only for the full flame mode. We extract source and background spectra from a circular region of $0^\prime.5$ radius and an annulus of $0^\prime.6$ inner and $1^\prime.1$ outer radii for all observations, respectively. Other processes, such as creating the Redistribution Matrix Files (RMFs) and the Auxiliary Response Files (ARFs), are performed in the same way as for the data reduction for J1856. All the spectra are binned in order to have more than 30 counts per bin using grppha.

\subsection{Examining the long-term stability}
\begin{figure}
 \begin{center}
  \includegraphics[width=12cm]{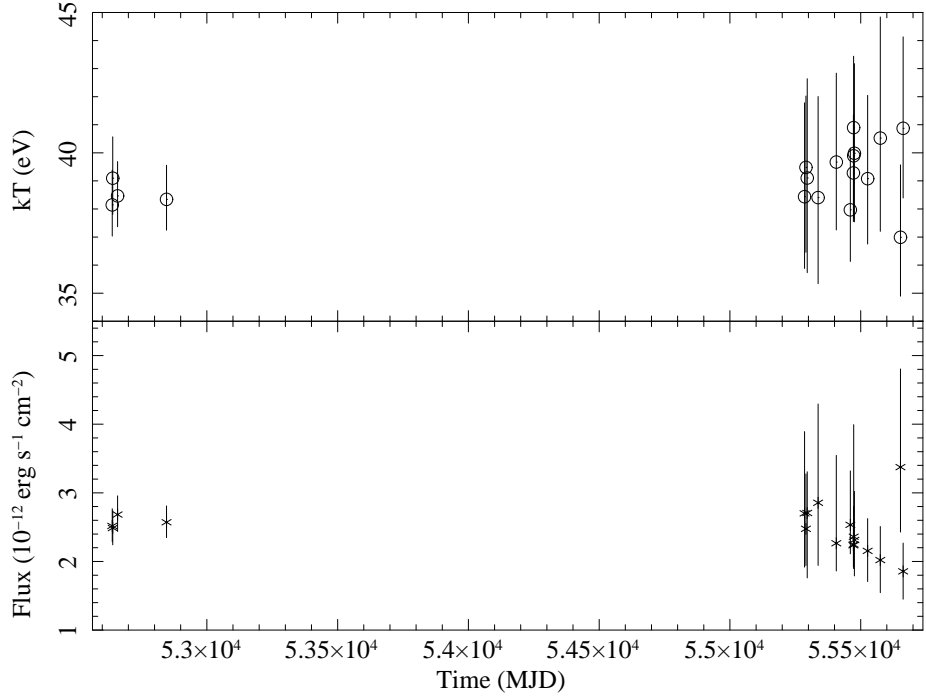} 
 \end{center}
\caption{Blackbody temperature (top panel) and unabsorbed bolometric flux (bottom) observed with XMM-Newton EPIC-pn for RX J0420.0$-$5022. The column density of interstellar matter $N_H$, the line center $E_{\rm line}$ and the line width $\sigma_{\rm line}$ are fixed as in table \ref{tab:ktevol_fixp}. The blackbody temperature $kT$, its normalization and the line normalization are left as free parameters.}\label{fig:j0420_ktflevol}
\end{figure}

\begin{figure}
 \begin{center}
  \includegraphics[width=12cm]{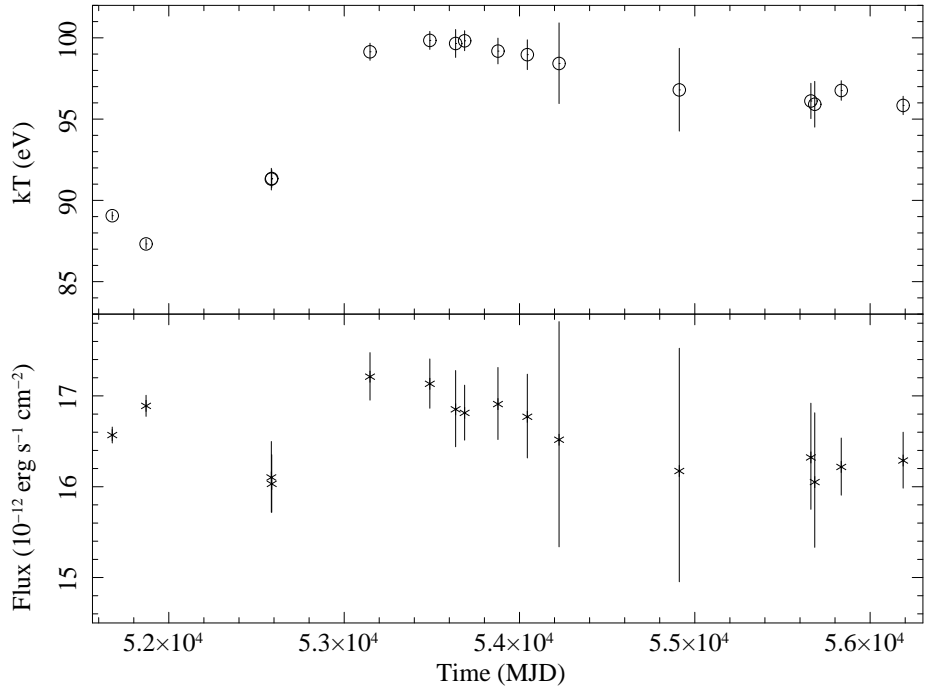} 
 \end{center}
\caption{Blackbody temperature (top panel) and unabsorbed bolometric flux (bottom) observed with XMM-Newton EPIC-pn for RX J0720.4$-$3125. Parameter condition is the same with figure \ref{fig:j0420_ktflevol}.}\label{fig:j0720_ktflevol}
\end{figure}

\begin{figure}
 \begin{center}
  \includegraphics[width=12cm]{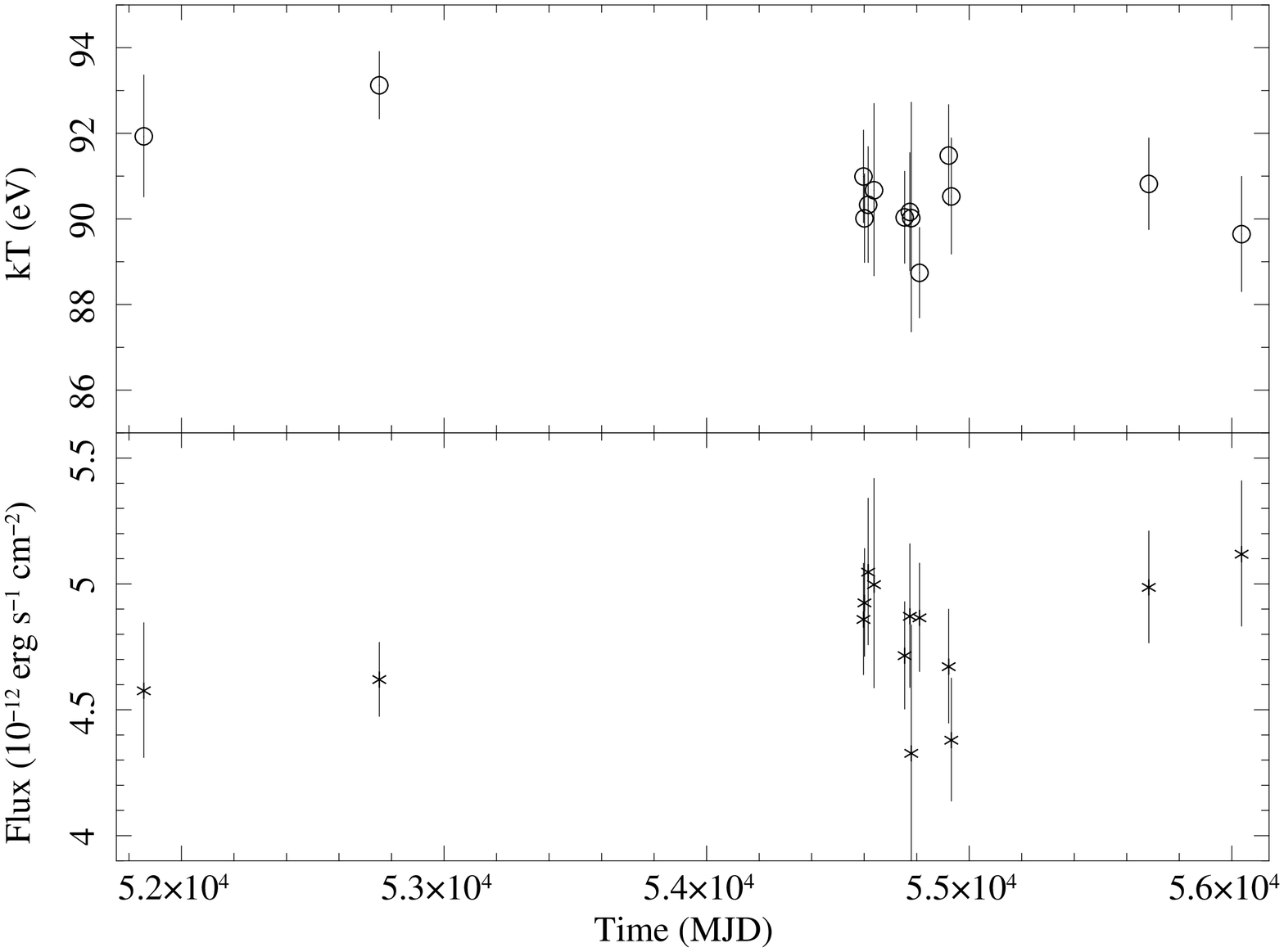} 
 \end{center}
\caption{Blackbody temperature (top panel) and unabsorbed bolometric flux (bottom) observed with XMM-Newton EPIC-pn for RX J0806.4$-$4123. Parameter settings are the same as those in figure \ref{fig:j0420_ktflevol}.}\label{fig:j0806_ktflevol}
\end{figure}

\begin{figure}
 \begin{center}
  \includegraphics[width=12cm]{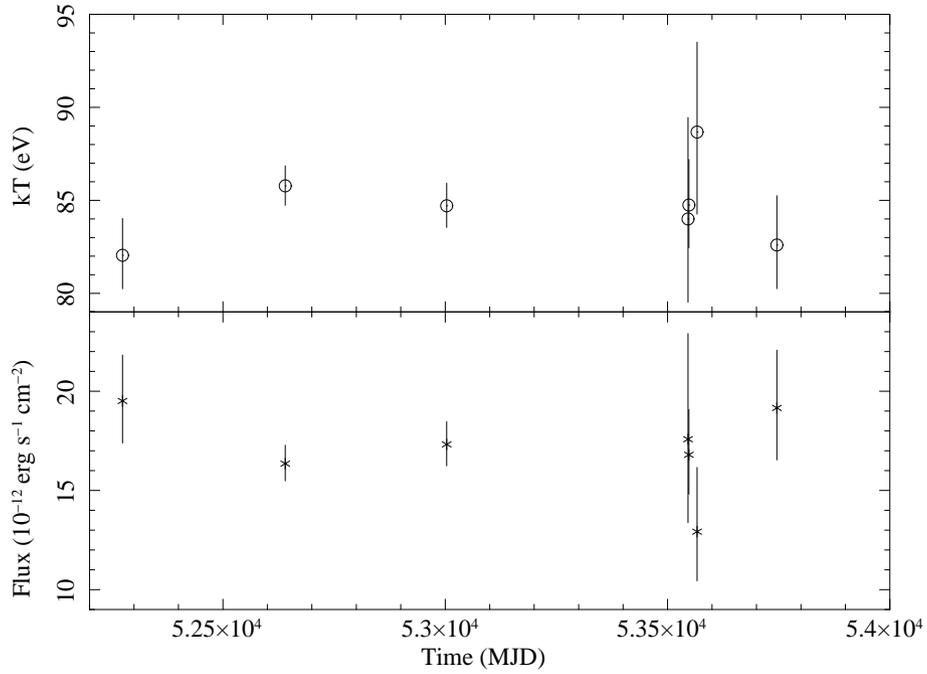} 
 \end{center}
\caption{Blackbody temperature (top panel) and unabsorbed bolometric flux (bottom) observed with XMM-Newton EPIC-pn for 1RXS J130848.6$+$212708 (RBS1223). Parameter settings are the same as those in figure \ref{fig:j0420_ktflevol}.}\label{fig:rbs1223_ktflevol}
\end{figure}

\begin{figure}
 \begin{center}
  \includegraphics[width=12cm]{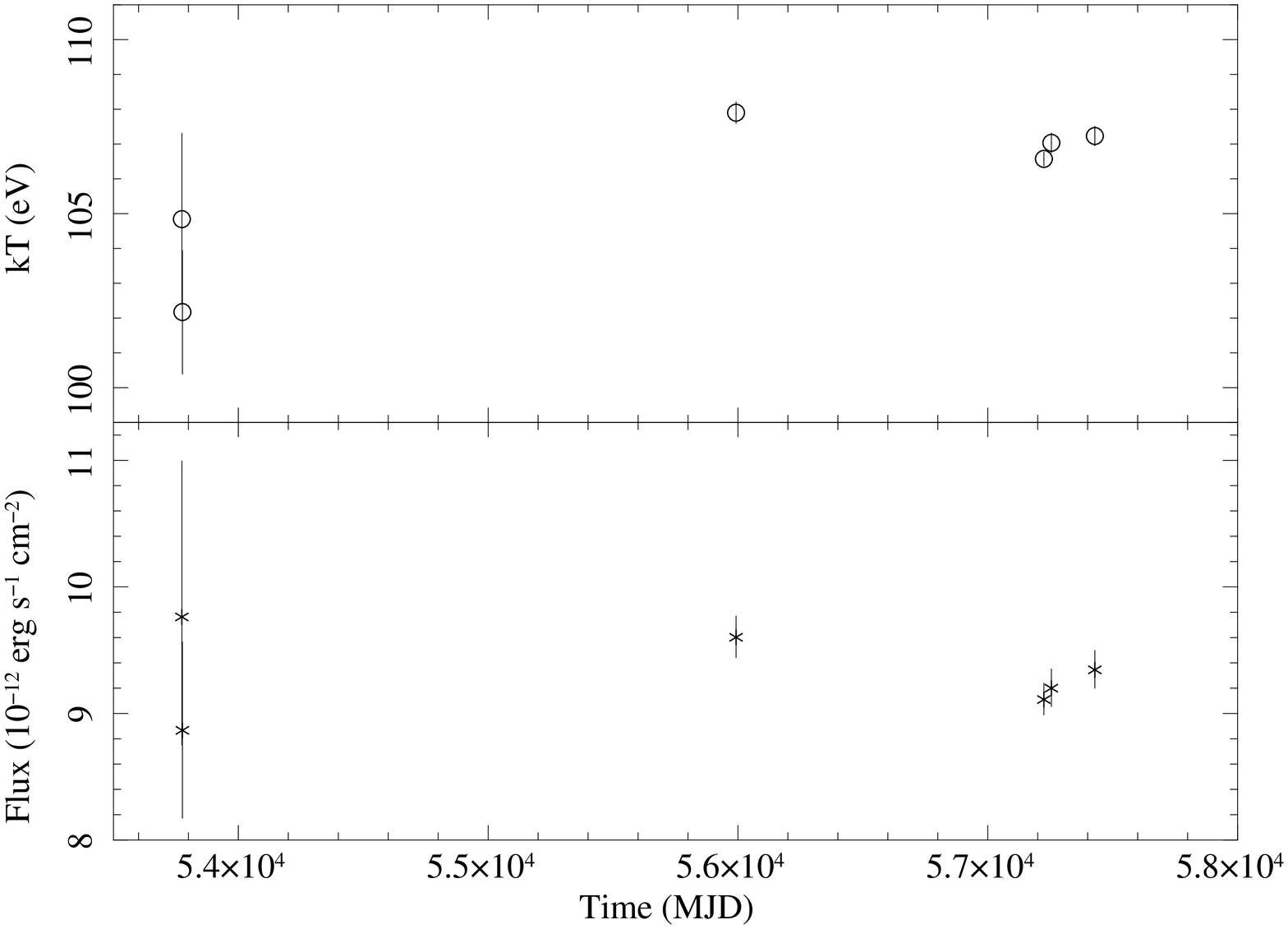} 
 \end{center}
\caption{Blackbody temperature (top panel) and unabsorbed bolometric flux (bottom) observed with XMM-Newton EPIC-pn for RX J1605.3$+$3249. Parameter settings are the same as those in figure \ref{fig:j0420_ktflevol}.}\label{fig:j1605_ktflevol}
\end{figure}

\begin{figure}
 \begin{center}
  \includegraphics[width=12cm]{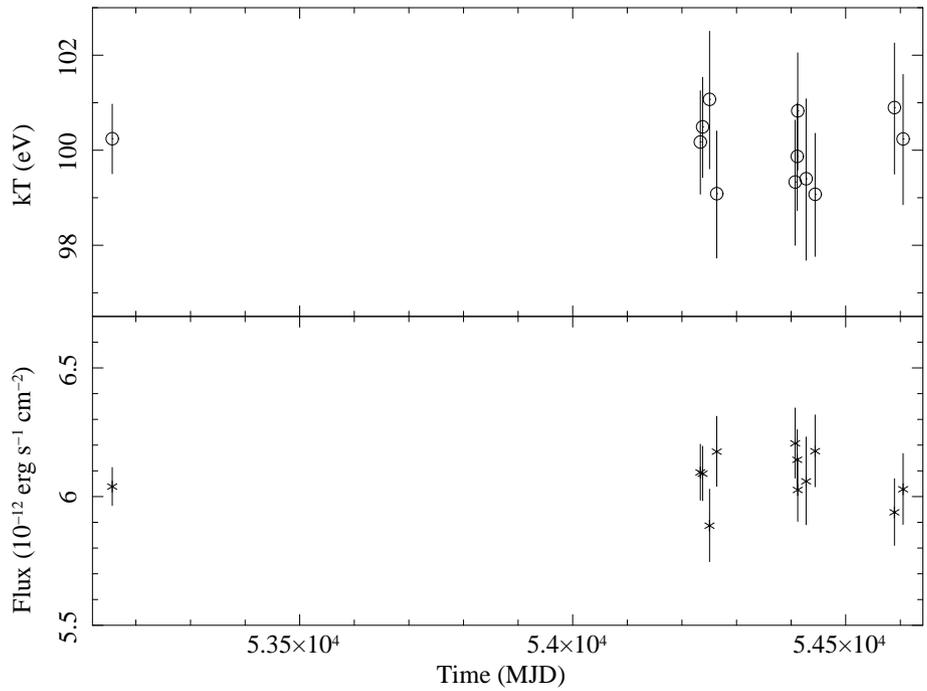} 
 \end{center}
\caption{Blackbody temperature (top panel) and unabsorbed bolometric flux (bottom) observed with XMM-Newton EPIC-pn for 1RXS J214303.7$+$065419 (RBS1774). Parameter settings are the same as those in figure \ref{fig:j0420_ktflevol}.}\label{fig:rbs1774_ktflevol}
\end{figure}

In Y17, we used the merged spectra of J1856. Being similar to the case of J1856, any single observation of any of the M7 does not provide enough photons for our detailed spectral analysis at around 1 keV, where the keV-excess emission is observed; therefore, to create the merged spectra for each source is essential to maximize the statistics. In merging the spectra, all the spectra should be similar to one another, which was indeed the case for J1856. We examine the long-term stability of the other six sources of the M7.
We fit all the spectra with a single temperature blackbody with a Gaussian absorption line model, or \verb'phabs*(bbodyrad+gauss)' in \verb'XSPEC ver12.9.1m'. Since J1605 shows two absorption lines, we add one more line for it. Note that the column density of interstellar matter $N_H$, the line center $E_{\rm line}$ and the line width $\sigma_{\rm line}$ are fixed as in table \ref{tab:ktevol_fixp}. All the other parameters, blackbody temperature $kT$, its normalization, and the line normalization are left as free parameters. Figure \ref{fig:j0420_ktflevol} to \ref{fig:rbs1774_ktflevol} shows the temperature and flux variation of the six sources, J0420, J0720, J0806, RBS1223, J1605 and RBS1774. We find that J0420, J0806, RBS1223, J1605 and RBS1774 do not show significant variability with 90\% confidence range, less than 5 eV in the temperature and $\sim 5 \times 10^{-20}$ erg s$^{-1}$ cm$^{-2}$ in the flux, whereas J0720 showed a significant variability as reported (Hohle et al. 2012). However, the temperature and flux of J0720 was stable during the period from 53,100 to 54,200 MJD. We thus merge all the spectra for each of the stable five sources and those in the stable period for J0720. 

\subsection{Discovery of the keV-excess}
\begin{figure}
 \begin{center}
  \includegraphics[width=12cm]{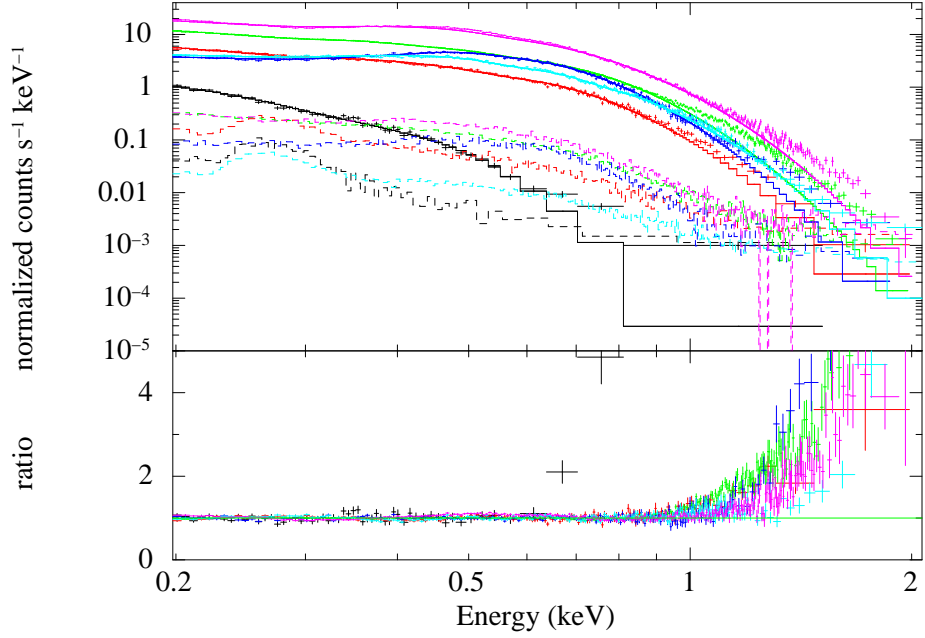} 
 \end{center}
\caption{X-ray spectra of six XINSs, RX J0420.0$-$5022 (black), RX J0720.4$-$3125 (magenta), RX J0806.4$-$4123 (red),  1RXS J130848.6+212707 (RBS1223; blue), RX J1605.3+3249 (green), and 1RXS J214303.7+065419 (RBS1774; light blue). We merge all the data of EPIC-pn for each sources except J0720. For J0720, we use the data in the stable period (53,100 to 54,200 MJD; see figure \ref{fig:j0720_ktflevol} and section 2.2). The solid lines are single temperature blackbody model with Gaussian absorption. Fitting parameters are shown in table \ref{tab:ktevol_fixp} and \ref{tab:1tbb}. The dashed lines are subtracted background spectra of each sources.}
\label{fig:1tbb}
\end{figure}

We fit each of the merged spectra with a single temperature blackbody and Gaussian absorption model, fixing $N_H$ and $E_{\rm line}$ as in table \ref{tab:ktevol_fixp}. Figure \ref{fig:1tbb} and table \ref{tab:1tbb} shows the results of all the six sources, together with J1856 adopted from Y17. A significant excess from the model is clearly visible for all the sources. Chi-square tests for these fits yield a chance probability of $10^{-5}$ or smaller for all the sources and confirm the excess to be significant. This result indicates the keV-excess is ubiquitous for the M7, unless it is not an artifacts or systematic error.

\section{Examining the keV-excess}
In Y17 for J1856, we defined the excess fraction, $f_{ex} = (c_{\rm obs} - c_{\rm mod}) / c_{\rm mod}$, where $c_{\rm obs}$ is the observed (background-subtracted) count rate and $c_{\rm mod}$ is the count rate expected from the baseline single-temperature blackbody model, in order to evaluate the excess emission, and we set the energy band to calculate $f_{\rm ex}$ to be 0.8 - 1.2 keV, where the excess is significant. We now define the energy range of $f_{\rm ex}$ for each source as listed in table \ref{tab:f_ex}. The range is 0.4 keV width and the centroid is proportional to the baseline blackbody temperature of the source with respect to that of J1856 (62.8 eV). For example, J0420 exhibits $kT = 42.8$ eV and thus the centroid is calculated to be $(42.8 / 62.8) \times 1.0$ keV $=$ 0.77 keV. All of the six sources show two times or more larger $f_{\rm ex}$ and lower X-ray flux than those of J1856, suggesting that the three possible systematic errors examined in Y17 (background fluctuation, pile-up, and confusion sources) should be less significant than those for J1856. The spatial background fluctuation of EPIC-pn is smaller than 18\% at 1 keV (Katayama et al. 2004), whereas our examination performed in Y17 showed the upper limit of 38\% with a 90\% confidence level for the area of $0^\prime.6$ – $1^\prime.1$ annulus. Employing the latter, we evaluate $\Delta b = 0.38 b$ for the background fluctuation, where $b$ is the background count rate. The systematic error for $f_{\rm ex}$ is given by $\Delta b / c_{\rm mod}$, where $c_{\rm mod}$ is the count rate expected from the baseline blackbody model. Table \ref{tab:f_ex} shows that $\Delta b / c_{\rm mod}$ is much smaller than $f_{\rm ex}$, even for J0420, the faintest among the M7. We also estimated the pile-up count rate $c = (4.56\pm0.03) \times 10^{-6}$ s$^{-1}$ and $c/c_{\rm mod} = (2.53\pm0.02)\times 10^{-2}$ for J1856. Since J1856 is the brightest, the count rates of the other six sources are smaller than that of J1856. Hence, the pile-up effect is less significant for the six sources, indicating that the pile-up cannot explain the excess emission observed. These six sources exhibit an excess flux of $c_{\rm obs} - c_{\rm mod} > 1\times10^{-3}$ s$^{-1}$ in their respectable energy bands of a 0.4 keV width. Converting this EPIC-pn count rate to the energy flux by using \verb'PIMMS', where a power-law spectrum with a photon index $\Gamma = 2$ is assumed, we obtain the X-ray flux $> 2 \times 10^{-14}$ erg s$^{-1}$ cm$^{-2}$ in an energy band of 0.1 - 10 keV. If the excess emission originated from contaminating sources, they should be detected and spatially resolved with high-resolution X-ray observations by the Chandra satellite. No such sources are detected within 0$^\prime$.5 circle of any of the targets of the M7. Therefore, we conclude that the excess emission is intrinsic to the M7 sources.

\section{Spectral fitting including the keV-excess}
We try to approximate the excess emission by fitting each X-ray spectrum of the M7 with, in addition to the already employed 1BB model, two separate models of another blackbody (collectively referred to as 2BB model) and a power-law (BB+PL model), allowing all parameters to be free and including a single Gaussian absorption in either case. We add another blackbody component with $kT_{\rm opt} = 32.3$ eV for J1856, which is responsible for the optical emission (Beuermann, Burwitz and Rauch 2006). 

\begin{figure}
 \begin{center}
  \includegraphics[width=12cm]{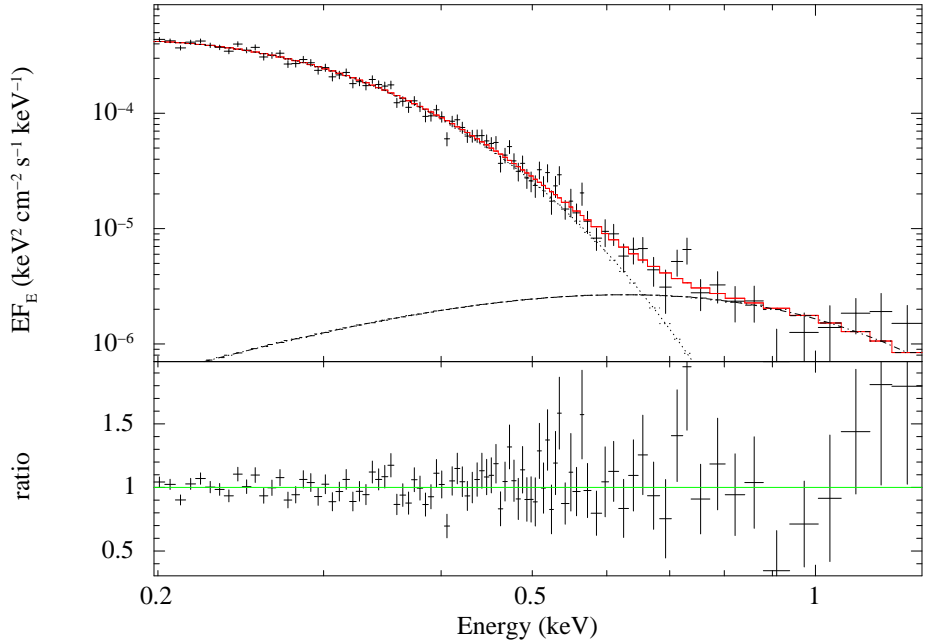} 
 \end{center}
\caption{Fit with the two-temperature blackbody model of J0420. The red solid line is the model, while the dotted line is the lower temperature component and dot-dashed the higher.}\label{fig:j0420_2bb}
\end{figure}

\begin{figure}
 \begin{center}
  \includegraphics[width=12cm]{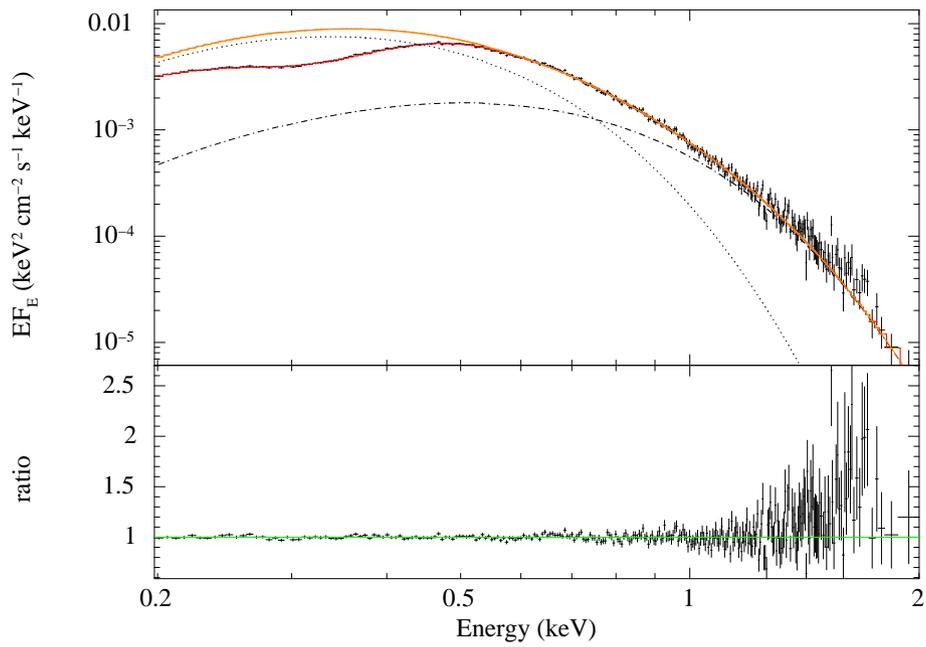} 
 \end{center}
\caption{Fit with the two-temperature blackbody model with gaussian absorption of J0720. The orange solid line correspond to the model without absorption.}\label{fig:j0720_2bb}
\end{figure}

\begin{figure}
 \begin{center}
  \includegraphics[width=12cm]{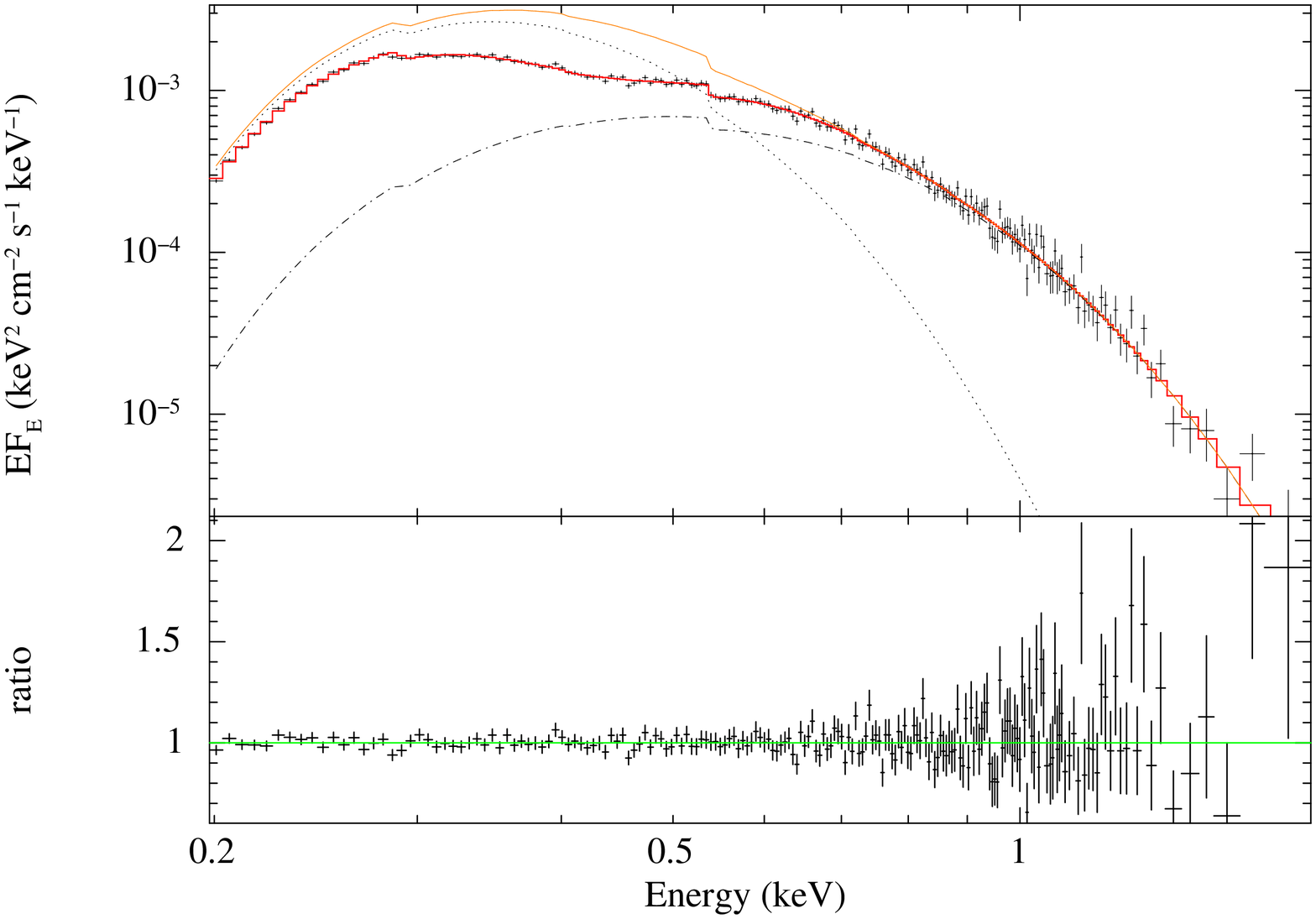} 
 \end{center}
\caption{Fit with the two-temperature blackbody model of J0806. Line indication is the same as that in figure \ref{fig:j0720_2bb}.}\label{fig:j0806_2bb}
\end{figure}

\begin{figure}
 \begin{center}
  \includegraphics[width=12cm]{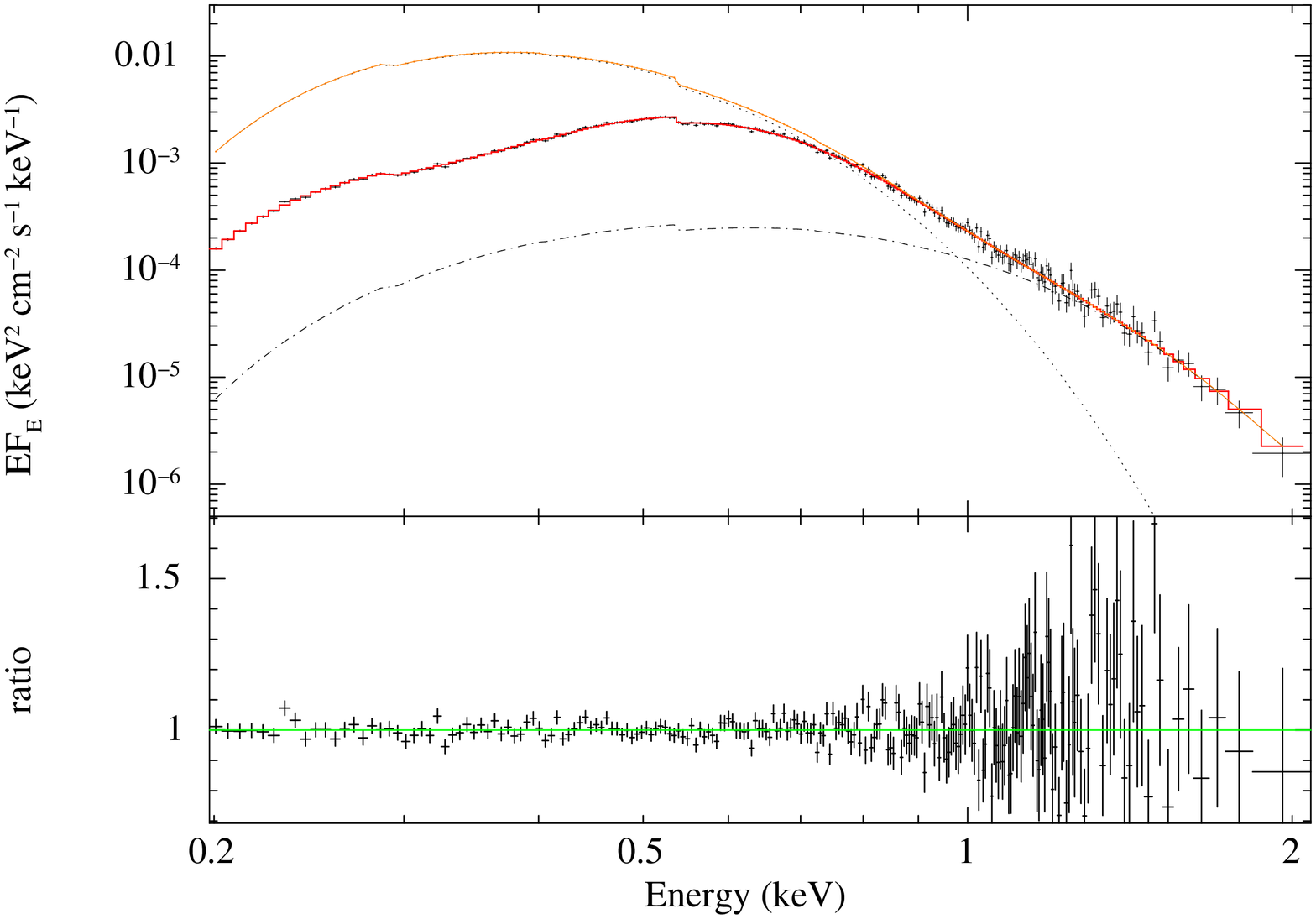} 
 \end{center}
\caption{Fit with the two-temperature blackbody model of RBS1223. Line indication is the same as that in figure \ref{fig:j0720_2bb}.}\label{fig:rbs1223_2bb}
\end{figure}

\begin{figure}
 \begin{center}
  \includegraphics[width=12cm]{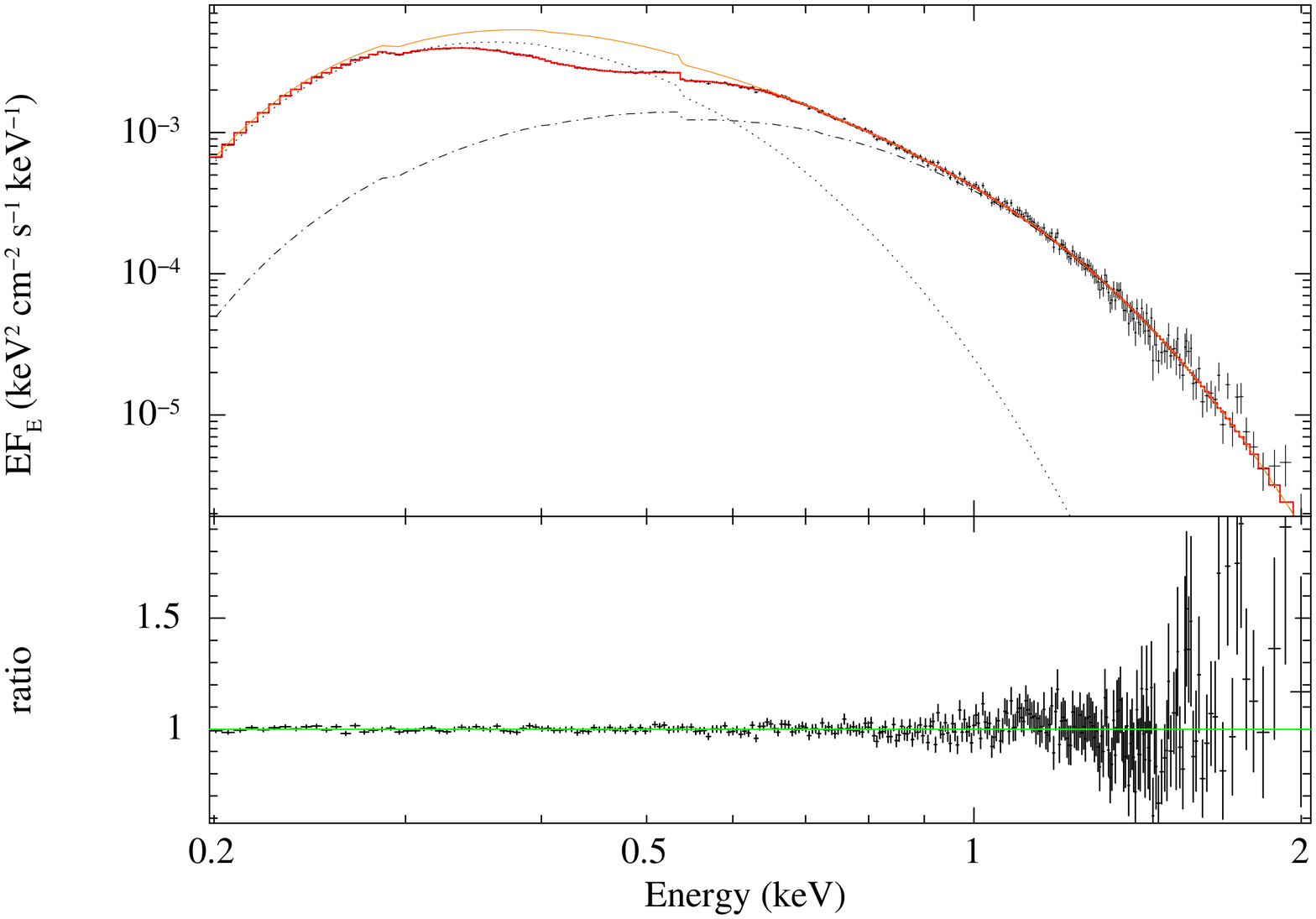} 
 \end{center}
\caption{Fit with the two-temperature blackbody model of J1605. Line indication is the same as that in figure \ref{fig:j0720_2bb}.}\label{fig:j1605_2bb}
\end{figure}

\begin{figure}
 \begin{center}
  \includegraphics[width=12cm]{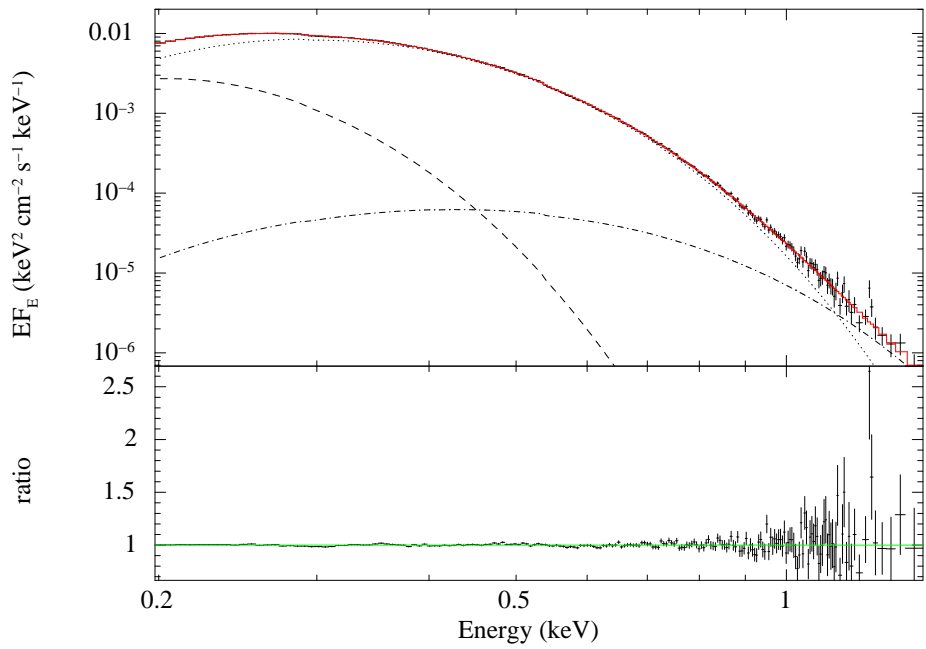} 
 \end{center}
\caption{Fit with the two-temperature blackbody model of J1856. The dashed line corresponds to lowest component with $kT_{\rm s} = 32.3$ eV.}\label{fig:j1856_3bb}
\end{figure}

\begin{figure}
 \begin{center}
  \includegraphics[width=12cm]{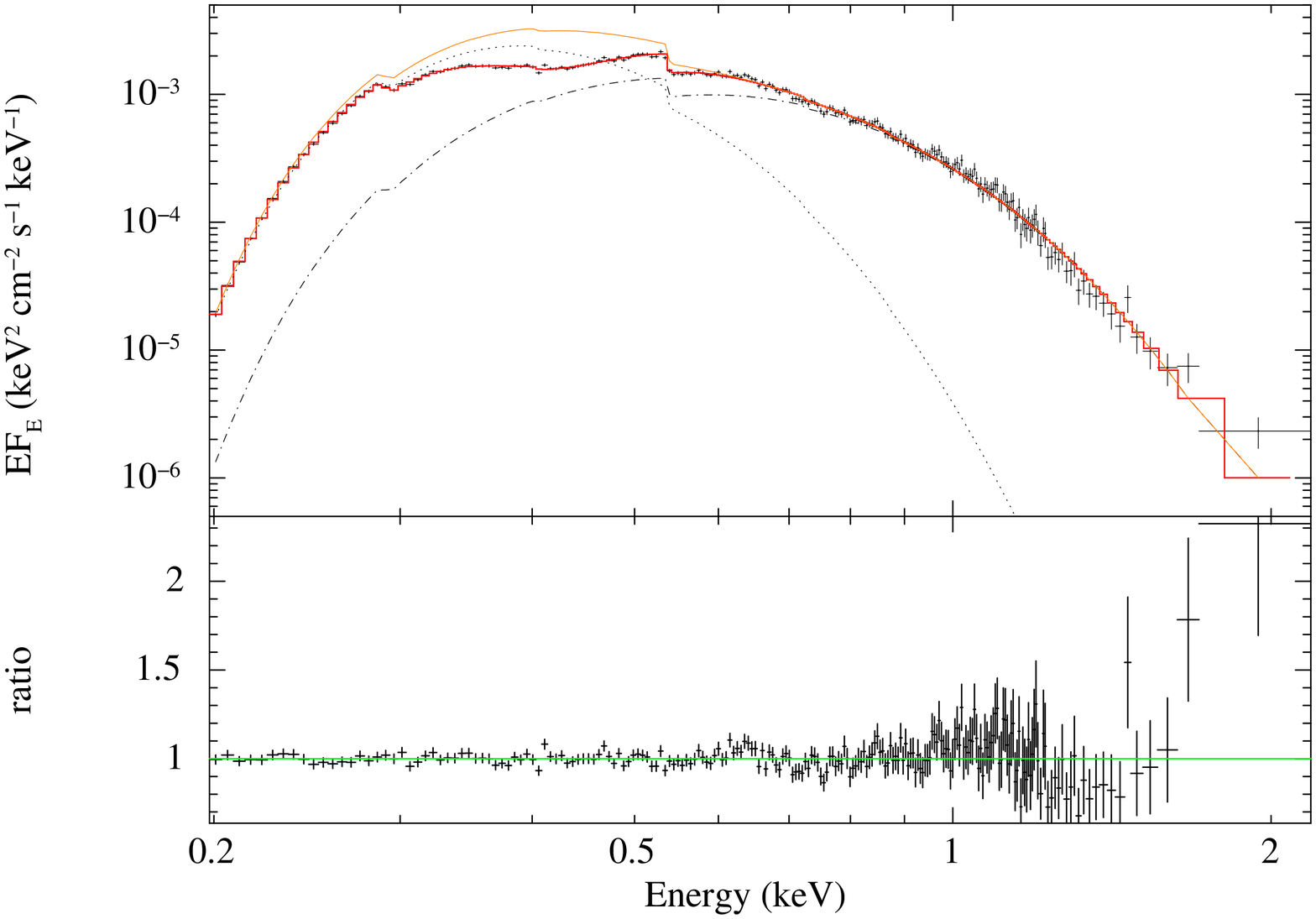} 
 \end{center}
\caption{Fit with the two-temperature blackbody model of RBS1774. Line indication is the same as that in figure \ref{fig:j0720_2bb}.}\label{fig:rbs1774_2bb}
\end{figure}

The 2BB model is found to significantly reduce the residuals from the 1BB model case for all the sources with temperatures of the cool ($kT_{\rm c}$) and hot ($kT_{\rm h}$) components of $kT_{\rm c} < 80$ eV and $kT_{\rm h} > 100$ eV, respectively (see table \ref{tab:2bb_par}). The spectra of all sources are shown in figures \ref{fig:j0420_2bb} to \ref{fig:rbs1774_2bb}. We obtain only the upper limits of interstellar absorption in J0420 and J0720. The absorption-line feature is not required for J0420 and J1856, while the other sources need broad absorption lines. We can estimate the blackbody radii of each component, using the normalization and the distance of each object. For the two sources of the M7, J1856 and J0720, their parallax distances are known(Walter et al. 2010 for J1856; Kaplan, van Kerkwijk and Anderson 2007 for J0720). The distances of the other sources of the M7 are determined from their interstellar X-ray absorptions (Posselt et al 2007), except for RBS1223, for which the generally accepted value of 500 pc is used because its high galactic latitude renders the X-ray-absorption method unreliable. Table \ref{tab:dist_r} lists the obtained blackbody radii for both the two components, $R_{\rm c}$ and $R_{\rm h}$. 

J1856 shows slightly different result from our previous work (Y17) because of the fitting condition. Pires et al. (2014) tested the 2BB model with a single absorption line for J1605. Their result is slightly different from ours, especially in $kT_{\rm c}$ (77 eV in their result and 65 eV in ours). For other six sources, this study is the first report of the existence of high-temperature emission in XINSs.

Table \ref{tab:bbpl_par} shows the results with the BB+PL model. For three sources of the M7, J0420, J0806, and J1856, this model provides similar $\chi^2$ as for the 2BB model, whereas for the other four sources this model is rejected on the basis of the $\chi^2$ values of the fitting.

\section{Discussion}

\subsection{Comparison of the spectral parameters between XINSs and magnetars}
\begin{figure}
 \begin{center}
  \includegraphics[width=12cm]{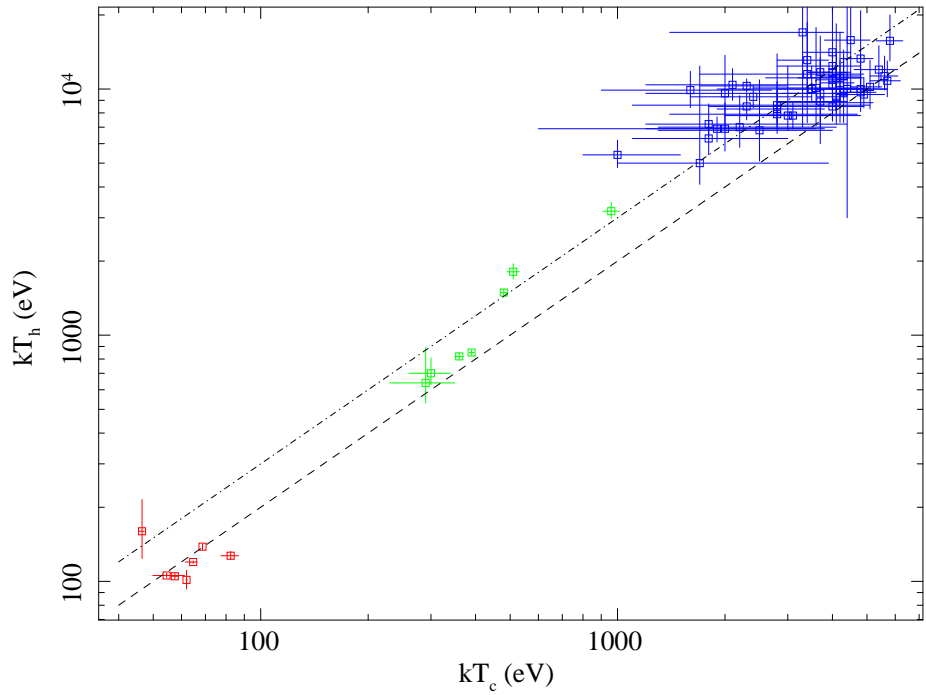} 
 \end{center}
\caption{Relation between the lower ($kT_{\rm c}$) and the higher ($kT_{\rm h}$) blackbody temperature for the M7 (red), magnetars in the quiescent (green; Nakagawa et al. 2009) and magnetars in the burst (blue; Nakagawa et al. 2007). Dashed and dot-dashed line correspond to $kT_{\rm h} / kT_{\rm c} = 2$ and $3$, respectively.}\label{fig:tdist}
\end{figure}

The spectra of magnetars in the soft X-ray band are reproduced with a 2BB model in both the burst and the persistent emission. It is reasonable to compare the spectra of XINSs and magnetars which are universally reproduced with the same model. Therefore, we hereafter focus on the 2BB model, which can reproduce six of the M7 spectra, and 3BB model, which includes the optical component, for J1856. We characterize the cool and hot components in the 2BB model by comparing the spectral characteristics of XINSs with those of magnetars. The blackbody emission is characterized simply with two parameters, temperature ($kT$) and emission radius ($R$). Figure \ref{fig:kt-r} shows the cool and hot temperatures ($kT_{\rm c}$ and $kT_{\rm h}$) versus their radii ($R_{\rm c}$ and $R_{\rm h}$) of each blackbody component of the M7, as well as those of magnetars in the quiescent and burst phases taken from Nakagawa et al. (2007, 2009). The data points of the M7 are found to show a similar trend to those of the quiescent emission of magnetars, i.e., $R_{\rm c} \sim 10$ km and $R_{\rm h} \sim 1$ km, except for those of the softest two of the M7, J0420 and J1856. Note that the uncertainty in distance is not taken into account. The error bars indicate only the statistical errors in the spectral fitting. We consider that this may be the reason why some XINSs and quiescent magnetars show significantly larger $R_{\rm c}$ than 12 km, a typical NS radius. In contrast, the temperatures are clearly separated between the M7 and magnetars by one order of magnitude. However, we find that the temperatures of the M7 and the magnetars in the both states universally satisfy the relation of $kT_{\rm h} / kT_{\rm c} \sim$ 2 - 3 (figure \ref{fig:tdist}). Although the temperatures are significantly different, the sizes of emission areas and the temperature ratio are very similar between the two families. These scaling relations strongly support the ``worn-out'' hypothesis. Furthermore, we find that the temperatures are clearly separated between XINSs and magnetars, whereas they are placed adjacently in the $P$ - $\dot{P}$ diagram. We discuss this in section 5.3.

\subsection{Speculating the evolution scenario with high-temperature component}
\begin{figure}
 \begin{center}
  \includegraphics[width=12cm]{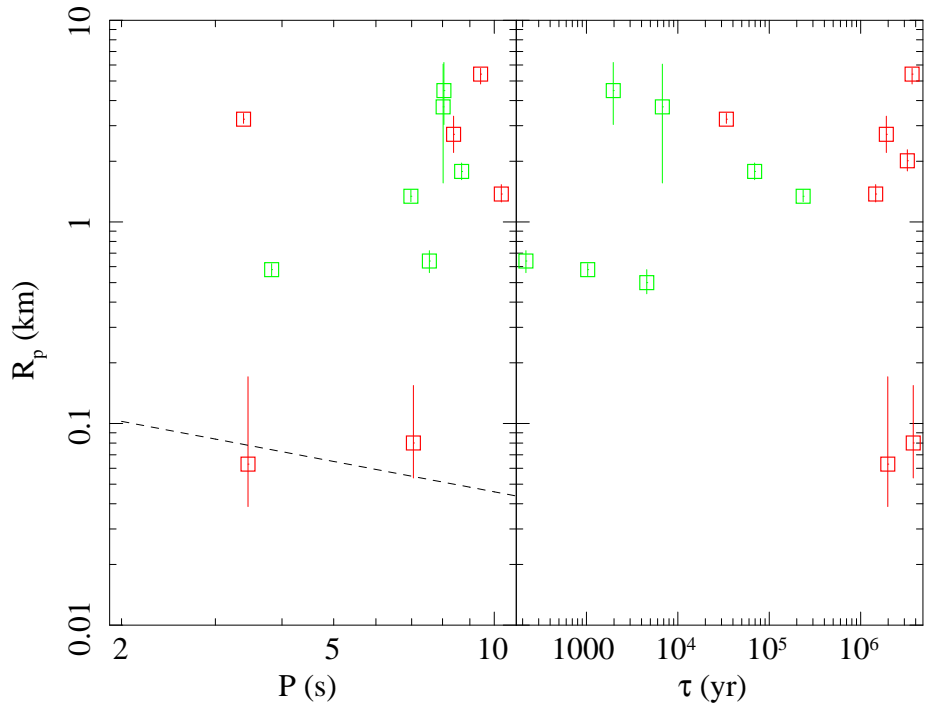} 
 \end{center}
\caption{$left$ $panel$: Relation between period ($P$) and the polar cap radius ($R_{\rm p} = 2 R_{\rm h}$) for the M7 (red) and magnetars in the quiescent (green). The dashed line shows the ``canonical model'', $R_{\rm pc} = 0.145 P^{-0.5}$. Most of the sources show significantly larger radii than the
model, although we should note that uncertainties in the distance and viewing angle can affect $R_{\rm p}$ by up to a factor of 2. The softest two of the M7, J0420 and J1856, exhibit radii consistent with those expected from the ``canonical model''. $right$ $panel$: The same plot, but plotted for the characteristic age $\tau$ (yr).}\label{fig:p-rp_tau-rp}
\end{figure}

The cool and hot components in the model are usually considered to originate from a large area of the NS surface and from the polar caps, respectively. The latter is thus considered to be responsible for the X-ray pulsation. The observed energy dependent pulsed fraction supports this model (e.g. Caraveo et al. 2004). On the basis of this hypothesis, we can estimate the polar cap radius $R_{\rm p}$. Assuming the emission area of the hot component is equal to the dual polar cap area whose shape is flat circle, we obtain a relation $4 \pi R_{\rm h}^2  = 2 \times \pi R_{\rm p}^2$, and the polar cap radius $R_{\rm p} = 2 R_{\rm h}$. According to the ``canonical model'' proposed by Goldreich and Julian (1969), the polar cap is defined as the locus of poloidal magnetic field lines penetrating a light cylinder, along which charged particles accrete onto the NS surface. We calculate the polar cap radius of the canonical model $R_{\rm pc} = 0.145 P^{-0.5}$ km, where $P$ (s) is a rotation period. Comparing $R_{\rm p}$ and $R_{\rm pc}$, we can examine whether the hot component corresponds to the “canonical” polar cap or not. Figure \ref{fig:p-rp_tau-rp} left panel shows the relation between $P$ and $R_{\rm p}$ for the M7 and the magnetars in the quiescent. Almost all the sources have larger radii than the canonical model irrespective to their family. Only two of the M7, J0420 and J1856, which are the softest sources with the weakest magnetic field among the M7 and magnetars, have $R_{\rm p}$ comparable with those expected from the canonical model. Note that we assume the viewing angle of the polar cap to be that of face-on case. The uncertainty of the viewing angle affects the results at most a factor of 2, which is not significant, given that most of the derived results are larger by more than an order. 

It is suggested that NSs have more complex magnetic field than a simple poloidal, a major part of the excess magnetic field is a toroidal (e.g., \"{O}zel 2013) one inside the star. Release of the internal magnetic energy via several ways (e.g. crust cracking in the NSs; Thompson and Duncan 1995, hydrodynamic deformation; Corsi and Owen 2011) can be the source of bursts observed on magnetars. Nakagawa et al. (2009) suggested that the emission mechanisms might be common in between the quiescent and burst phases, considering the parameter correlation. This is analogous to microflares and ordinary flares of the Sun. We then speculate a scenario; thermal energy from decaying complex magnetic field might heat up a larger area of the surface than that of the canonical polar cap. Only when the troidal field decays and when the global field equilibrates, we observe the hot component from the canonical polar cap, as in the cases of J0420 and J1856. A theoretical study will support this scenario. Perna et al. (2013) modeled a surface temperature distribution for magnetized NS considering both poloidal and troidal magnetic field. They concluded that the pure poloidal field makes nearly isothermal NS surface, while existence of toroidal field makes hot spots and extended cooler region on the surface. In this scenario, the age of NSs should correlate with $R_{\rm p}$. However, according to the characteristic age ($\tau$), the two NSs with the weakest magnetic field and smaller $R_{\rm p}$ are not older than the other members of the M7, while they are significantly older than magnetars (the right panel of figure \ref{fig:p-rp_tau-rp}). Note that we assume that the magnetic field of a NS is constant during its life in deriving $\tau$ from the NS rotation. Considering the internal structure of NSs, this assumption may not be appropriate because the magnetic field should decay, even though the time scale depends on the evolution model. The true age of NSs may differ from $\tau$. This could be a reason why we do not find the correlation between $\tau$ and $R_{\rm p}$. 

\subsection{Unification and missing link between XINS and Magnetars}
As mentioned in the introduction, the ``worn-out'' magnetar hypothesis has been considered from the $P$ - $\dot{P}$ diagram and from the primarily temperature of their thermal emission.

Recently, it is suggested that some XINSs actually have a complex magnetic field. Borghese et al. (2015, 2017) found narrow phase-dependent absorption lines in J0720 and RBS1223 with an energy centroid of $\sim 750$ eV. Assuming the lines originate in resonant cyclotron absorption/scattering of protons that occured in a local strong magnetic field structure close to the stellar surface, they estimated the local field strength to $B \sim 2 \times 10^{14}$ G, about a factor of $\sim 5$ higher than the dipole magnetic fields. They suggested that these lines have similar property with a phase-dependent absorption feature detected in some low-field magnetars (Tiengo et al. 2013; Rodr\'{i}guez Castillo et al 2016), also supporting the ``worn-out'' hypothesis. Note, however, that their finding was so far limited to only two XINSs among the seven.

\begin{figure}
 \begin{center}
  \includegraphics[width=12cm]{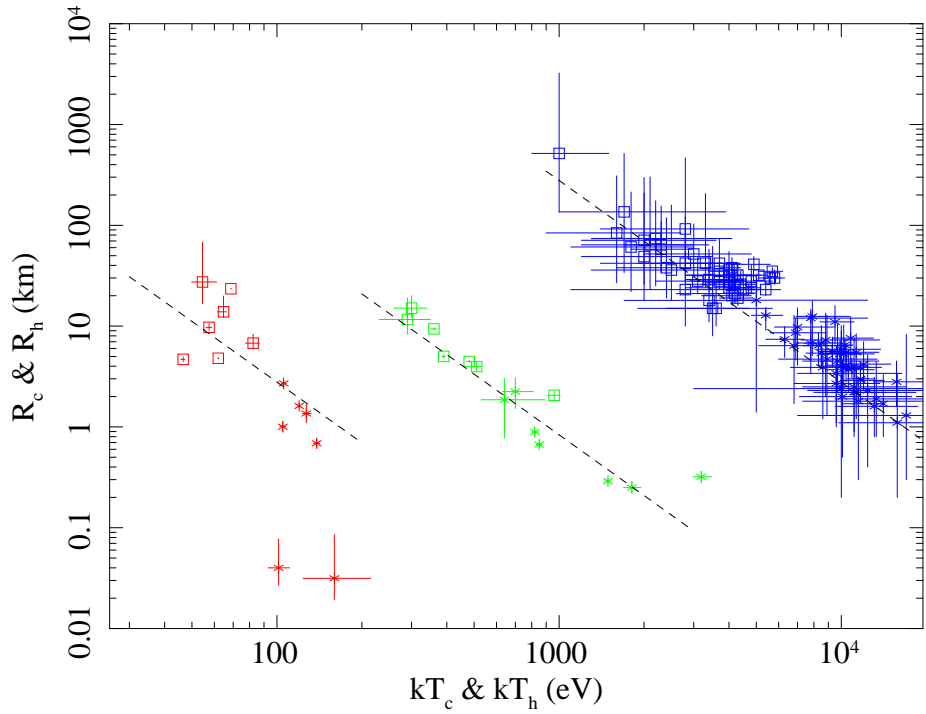} 
 \end{center}
\caption{Relation between temperatures of the 2BB model and corresponding radiation radii  for the M7 (red), magnetar persistent emission in the quiescent (green) and magnetar short bursts (blue). The cool ($kT_{\rm c}$) component are shown in square and the hot ($kT_{\rm h}$) is square. The dashed lines correspond to $L_{\rm X} = 10^{32}$, $10^{35}$, and $10^{40}$ erg s$^{-1}$ from left side, respectively. The magnetars' data are referred to Nakagawa et al. (2007, 2009). Note that the uncertainty in distance is not taken into account.
}\label{fig:kt-r}
\end{figure}

Our discovery of the keV-excess for all the  seven XINSs and its unification to the high temperature component of magnetars, as summarized in figure \ref{fig:kt-r}, further enhance the ``worn-out'' hypothesis of XINSs. 
Theoretical study also supports the hypothesis. Vigan\'{o} et al. (2013) established magneto-thermal evolution models for various families of isolated NSs. They suggested that magnetars and XINSs lie on the same model of evolution track.

We, however, consider the gap between XINSs and magnetars is as important as the unification of these two families. The separation is not clear in the $P$ - $\dot{P}$ diagram but impressively illustrated in our figure \ref{fig:kt-r}. If we refer the theoretical model for magneto-thermal evolution of isolated NS by Vigano et al. (2013), there should be isolated NSs with the thermal luminosity of $\sim 10^{34}$ erg s$^{-1}$ in the middle of magnetars and XINSs, though they did not explicitely mentioned. Recently, quiescent states of the transient magnetars are observed. (e.g. Enoto et al. 2017, Zelati et al. 2017). Their thermal X-ray luminosity is reported to be $10^{33}$ -- $10^{34}$ erg s$^{-1}$, an intermediate value between XINSs ($\sim 10^{32}$ erg s$^{-1}$) and bright quiescent magnetar ($\sim 10^{35}$ erg s$^{-1}$) employed by Nakagawa et al. (2009) in figure \ref{fig:kt-r}. Hence, the gap can be filled with the quiescent state emission of transient magnetars. Examination of their X-ray spectra with the 2BB model will be a good test for this hypothesis.

\section{Summary}
In our previous work, we presented that RX J1856.5-3754, the brightest source in the Magnificent Seven, exhibits a high-energy excess emission over the known two-temperature blackbody model around 1 keV. In this paper, we find the ``keV-excess'' from all the other objects in the Magnificent Seven (M7). The keV-excess is evaluated with the excess fraction, $\sim 16$\% for J1856 and $\sim 33$ -- $90$\% for the other six sources. We examine possible systematic errors for the six sources as well as for J1856 in Y17. However, none of them can explain the keV-excess. We thus conclude that the keV-excess is universal for the M7. We then try to approximate the keV-excess emission by fitting each X-ray spectrum of the M7 with two-temperature blackbody (2BB) and single-temperature blackbody with a power-law (BB+PL) model. The former can reproduce all the seven sources with temperatures $kT_c < 80$ eV for cool component and $kT_h > 100$ eV for hot component, whereas the latter is acceptable only for three sources.

The soft X-ray spectra of magnetars can be also reproduced with the 2BB model. Therefore we compare the spectra of XINSs and magnetars. We find that distribution of the two temperatures and their radii ($R_c$, $R_h$) are remarkably similar between XINSs and magnetars, while the temperatures are different with one order of magnitude. This similarity suggests that their origins of the emission are also the same. We thus state that this result is the strongest observational indication that supports the ``worn-out'' hypothesis. 

We first consider that the hot component is originate in polar caps of the NS surface. Comparing the observed polar cap radii $R_p$ derived from $R_h$ with the canonical polar cap model, we find that most of the sources of XINSs and magnetars show an order of magnitude larger radii than the canonical model, whereas the two XINSs with the lowest temperatures and the weakest magnetic fields show similar radii. We thus speculate that thermal energy from decaying local complex magnetic field in the star might heat up a larger area of the surface than that of the canonical polar cap. Only when the complex field decays and when the global field equilibrates, we observe the hot component from the canonical polar caps.

Our result arise a new question. The temperature is separated clearly between magnetars and XINSs, whereas they are placed adjacently in the $P$ - $\dot{P}$ diagram. This gap can be filled with the quiescent emission of the transient magnetars which show intermediate X-ray luminosities between XINSs and bright quiescent magnetars, if their X-ray spectra are universally reproduced with the 2BB model. We need further theoretical and observational studies to understand the relationship between XINSs and magnetars, or more ambitiously, to establish the Grand Unification of Neutron Stars (GUNS; Kaspi 2010).



\begin{ack}
This work is supported by Japan Society for the Promotion of Science (JSPS) KAKANHI Grant Numbers JP 15640356, 15H03641, 16H00949, 16K13787, 18H01256, and 18J20523.
\end{ack}






\newpage


\newpage



\begin{table}[htbp]
  \tbl{EPIC-pn observations of the six XINSs}{%
  \begin{tabular}{ccccccc}
\hline														
J0420	&	J0720	&	J0806	&	RBS1223	&	J1605	&	J1856	&	RBS1774	\\	\hline
Obs. ID	&	Obs. ID	&	Obs. ID	&	Obs. ID	&	Obs. ID	&	Obs. ID	&	Obs. ID	\\	\hline
0141750101	&	0124100101	&	0106260201	&	090010101	&	0302140101	&	0165971601	&	0201150101	\\	
0141751001	&	0132520301	&	0141750501	&	0157360101	&	0302140401	&	0165972001	&	0502040601	\\	
0141751101	&	0156960201	&	0552210201	&	0163560101	&	0671620101	&	0165972102	&	0502040701	\\	
0651470201	&	0156960401	&	0552210301	&	0305900201	&	0764460201	&	0165970301	&	0502040801	\\	
0651470301	&	0164560501	&	0552210401	&	0305900301	&	0764460401	&	0412600301	&	0502040901	\\	
0651470401	&	0300520201	&	0552210501	&	0305900401	&	0764460501	&	0412600601	&	0502041001	\\	
0651470501	&	0300520301	&	0552210601	&	0305900601	&		&	0412600801	&	0502041101	\\	
0651470601	&	0311590101	&	0552210901	&		&		&	0412601101	&	0502041201	\\	
0651470701	&	0400140301	&	0552211001	&		&		&	0412601501	&	0502041301	\\	
0651470801	&	0400140401	&	0552211101	&		&		&	0412602301	&	0502041401	\\	
0651470901	&	0502710201	&	0552211501	&		&		&	0727760101	&	0502041501	\\	
0651471001	&	0554510101	&	0552211601	&		&		&	0727750301	&	0502041801	\\	
0651471101	&	0650920101	&	0672980201	&		&		&	0727760501	&		\\	
0651471201	&	0670700201	&	0672980301	&		&		&		&		\\	
0651471301	&	0670700301	&		&		&		&		&		\\	
0651471401	&	0690070201	&		&		&		&		&		\\	
0651471501	&		&		&		&		&		&		\\	\hline
Total Exposure [ks]	&	Total Exposure [ks]	&	Total Exposure [ks]	&	Total Exposure [ks]	&	Total Exposure [ks]	&	Total Exposure [ks]	&	Total Exposure [ks]	\\	
135.02	&	205.75	&	74.27	&	60.52	&	247.69	&	392.72	&	87.74	\\	\hline
 \end{tabular}}\label{tab:pn_data}
\begin{tabnote}
Data of J1856 are adopted from Y17.
\end{tabnote}
\end{table}

\begin{table}[htbp]
  \tbl{Fixed parameters in temperature variation analysis}{%
  \begin{tabular}{cccc}
\hline  
Object	&	$N_{\rm H}$ [$10^{20}$ cm$^{-2}$]	&	$E_{\rm line}$ [eV]	&	$\sigma$ [eV] \\ \hline
J0420$^{[1]}$	&	2.02	&	329	&	70 \\
J0720$^{[2]}$	&	0.984	&	312	&	64 \\
J0806$^{[1]}$	&	1.12	&	460	&	70 \\
RBS1223$^{[3]}$	&	4.66	&	270	&	117 \\
J1605$^{[4]}$	&	0	&	441	&	128 \\
	&		&	790	&	101 \\
RBS1774$^{[5]}$	&	3.6	&	754	&	27 \\ \hline
    \end{tabular}}\label{tab:ktevol_fixp}
\begin{tabnote}
References are as follows: [1] Haberl et al. (2004), [2] Hohle et al. (2012), [3] Hambaryan et al. (2011), [4] Pires et al. (2014) for higher energy line, while lower is determined by simultaneous fitting, [5] Cropper et al. (2007).
\end{tabnote}
\end{table}

\begin{table}[htbp]
  \tbl{Parameters of single temperature blackbody model for merged spectra}{%
  \begin{tabular}{cccccc}
\hline  
Object	&	$kT$ [eV]	&	Norm. [10$^3$ (km / 10 kpc)$^2$]	&	$\sigma$ [eV]	&	EW [eV]	&	$\chi^2_r$ / dof \\ \hline
J0420	&	42.8	&	79.7	&	55.8	&	-48	&	2.95 / 57 \\
J0720	&	102.2	&	12.6	&	48.7	&	-45	&	3.15 / 316 \\
J0806	&	89.6	&	6.6	&	87.5	&	-58	&	2.66 / 176 \\
RBS1223	&	88.4	&	21.3	&	114.8	&	-128	&	2.18 / 188 \\
J1605	&	105.0	&	71.6	&	130.5	&	-90	&	3.51 / 251 \\
&&&112.7 & -73 & \\
RBS1774	&	104.6	&	3.9	&	33.0	&	-16	&	1.47 / 199 \\ \hline
 \end{tabular}}\label{tab:1tbb}
\begin{tabnote}
\end{tabnote}
\end{table}

\begin{table}[htbp]
  \tbl{Excess fraction for the six sources}{%
  \begin{tabular}{ccccccc}
\hline  
Object	&	Band [keV]	&	$c_{\rm obs}$ [$10^{-3}$ cts s$^{-1}$]	&	$c_{\rm mod}$  [$10^{-3}$ cts s$^{-1}$]		&	$f_{\rm ex}$	&	$b$  [$10^{-3}$ cts s$^{-1}$]	&	$\Delta b / c_{\rm mod}$ \\ \hline
J0420	&	0.57 - 0.97	&	2.47 $\pm$ 0.17	&	1.34	&	0.85 $\pm$ 0.15	&	0.86 $\pm$ 0.08	&	0.25 $\pm0.02$ \\
J0720	&	1.31 - 1.71	&	4.56 $\pm$ 0.24	&	3.44	&	0.33 $\pm$ 0.07	&	0.51 $\pm$ 0.07	&	0.056 $\pm$ 0.003 \\
J0806	&	1.17 - 1.57	&	5.02 $\pm$ 0.29	&	2.64	&	0.90 $\pm$ 0.12	&	1.04 $\pm$ 0.11	&	0.149 $\pm$ 0.009 \\
RBS1223	&	1.10 - 1.50	&	17.03 $\pm$0.54	&	9.99	&	0.70 $\pm$ 0.05	&	0.55 $\pm$ 0.09	&	0.0207 $\pm$ 0.0007 \\
J1605	&	1.21 - 1.61	&	15.61 $\pm$ 0.27	&	6.53	&	0.79 $\pm$ 0.01	&	0.79 $\pm$ 0.06	&	0.0458 $\pm$ 0.0008 \\
RBS1774	&	1.30 - 1.70	&	4.60 $\pm$ 0.24	&	3.20	&	0.44 $\pm$ 0.07	&	0.36 $\pm$ 0.06	&	0.042 $\pm$ 0.002 \\ \hline
J1856	&	0.80 - 1.20	&	20.70 $\pm$ 0.33	&	17.90	&	0.16 $\pm$ 0.02	&	0.25 $\pm$ 0.01	&	0.0044 $\pm$ 0.0001 \\ \hline
 \end{tabular}}\label{tab:f_ex}
\begin{tabnote}
The value for J1856 is from Y17.
\end{tabnote}
\end{table}

\begin{table}[htbp]
  \tbl{Fitting parameter of two temperature blackbody model}{%
  \begin{tabular}{cccccccccc}
\hline  
Object	&	$N_{\rm H}$ [$10^{20}$ cm$^{-2}$]	&	$kT_{\rm c}$ [eV]	&	Norm.$_{\rm c}^\ast$	&	$kT_{\rm h}$ [eV]	&	Norm.$_{\rm h}^\ast$	&	$E_{\rm line}$ [eV]	&	$\sigma$ [eV]	&	EW [eV]	&	$\chi^2_r$ / dof	\\ \hline
J0420	&	$< 0.21$	&	$46.5 ^{+0.7}_{-0.9}$	&	$18.4^{+2.3}_{-1.9}$	&	$160^{+55}_{-36}$	&	$(0.8^{+2.8}_{-0.6}) \times 10^{-3}$	&	--	&	--	&	--	&	1.22 / 85	\\
J0720	&	$< 0.64$	&	$82.4^{+4.3}_{-5.0}$	&	$35.4^{+16.0}_{-8.1}$	&	$127^{+5}_{-4}$	&	$1.4^{+0.7}_{-0.5}$	&	$254^{+25}_{-30}$	&	$97^{+13}_{-12}$	&	$-104^{+27}_{-21}$	&	1.37 / 354	\\
J0806	&	$3.82^{+0.23}_{-0.22}$	&	$57.5^{+1.3}_{-1.4}$	&	$151.2^{+1.3}_{-1.4}$	&	$104.9^{+2.7}_{-2.6}$	&	$1.62^{+0.04}_{-0.04}$	&	$241^{+11}_{-12}$	&	$125^{+4}_{-4}$	&	$-101.5^{+0.6}_{-1.4}$	&	0.96 / 194	\\
RBS1223	&	$3.14^{+0.23}_{-0.23}$	&	$68.7^{+0.1}_{-0.1}$	&	$220.6^{+2.0}_{-1.8}$	&	$138.3^{+3.7}_{-3.7}$	&	$0.19^{+0.04}_{-0.03}$	&	$390^{+6}_{-6}$	&	$183.6^{+1.6}_{-1.5}$	&	$-202.9^{+0.5}_{-0.3}$	&	1.06 / 229	\\
J1605	&	$3.39^{+0.74}_{-0.35}$	&	$64.7^{+1.6}_{-3.1}$	&	$126^{+110}_{-30}$	&	$119.9^{+1.2}_{-1.5}$	&	$1.72^{+0.23}_{-0.16}$	&	$353^{+19}_{-48}$	&	$96^{+15}_{-7}$	&	$-78.8^{+5.8}_{-4.3}$	&	1.00 / 282	\\
J1856	&	$1.00^{+0.03}_{-0.03}$	&	$62.0^{+0.2}_{-0.3}$	&	$164.7^{+4.9}_{-4.2}$	&	$101.3^{+9.4}_{-8.3}$	&	$0.13^{+0.23}_{-0.08}$	&	--	&	--	&	--	&	1.21 / 206	\\
RBS1774	&	$8.3^{+1.8}_{-2.1}$	&	$54.5^{+6.7}_{-4.6}$	&	$407^{+1196}_{-319}$	&	$105.6^{+2.2}_{-1.9}$	&	$3.96^{+5.3}_{-8.5}$	&	$326^{+56}_{-79}$	&	$87^{+23}_{-24}$	&	$-84^{+24}_{-7}$	&	1.13 / 218	\\ \hline
 \end{tabular}}\label{tab:2bb_par}
\begin{tabnote}
$^\ast$In unit of $10^3$ (km / 10 kpc)$^2$
\end{tabnote}
\end{table}

\begin{table}[htbp]
 \tbl{Blackbody radii of two-temperature model}{%
 \begin{tabular}{cccc}
\hline
Object		&	Distance [pc]	&	$R_{\rm c}$ [km]	&	$R_{\rm h}$ [km] \\ \hline
J0420	&	345	&	$4.68^{+0.30}_{-0.24}$	&	$0.03^{+0.05}_{-0.01}$	\\
J0720	&	360	&	$6.78^{+1.53}_{-0.78}$	&	$1.36^{+0.31}_{-0.26}$	\\
J0806	&	250	&	$9.72^{+0.82}_{-0.64}$	&	$1.00^{+0.14}_{-0.11}$	\\
RBS1223	&	500	&	$23.49^{+0.10}_{-0.10}$	&	$0.69^{+0.08}_{-0.06}$	\\
J1605	&	390	&	$13.86^{+6.01}_{-1.67}$	&	$1.62^{+0.11}_{-0.07}$	\\
J1856	&	123	&	$4.80^{+0.05}_{-0.04}$	&	$0.04^{+0.04}_{-0.01}$	\\
RBS1774	&	430	&	$27.4^{+40.3}_{-10.7}$	&	$2.71^{+0.18}_{-0.29}$	\\ \hline
 \end{tabular}}\label{tab:dist_r}
\begin{tabnote}
The distances are determined by parallax measurement for J0720 (Kaplan, Kerkwijk and Anderson 2007) and J1856 (Walter et al. 2010), X-ray absorption for J0420, J0806, J1605 and RBS1774 (Posselt et al. 2007). We assume 500 pc for RBS1223 as Kaplan and Kerkwijk (2009).
\end{tabnote}
\end{table}

\begin{table}[htbp]
  \tbl{Fitting parameter of single temperature blackbody with power-law model}{%
  \begin{tabular}{cccccccccc}
\hline  
Object	&	$N_{\rm H}$ [$10^{20}$ cm$^{-2}$]	&	$kT$ [eV]	&	Norm.$_{\rm bb}$ 	&	$\Gamma$	&	Norm.$_{\rm pl}$ 	&	$E_{\rm line}$ [eV]	&	$\sigma_{\rm line}$ [eV]	&	EW [eV]	&	$\chi^2_r$ / dof	\\	
	&		&		&	[10$^3$ (km / 10 kpc)$^2$]	&		&	[$10^{-5}$ s$^{-1}$ keV$^{-1}$ at 1 keV]	&		&		&		& \\ \hline	
J0420	&	$< 8.2$	&	$46.1$	&	$18.7$	&	$3.66$	&	$0.16$	&	--	&	--	&	--	&	1.17 / 86	\\	
J0720	&	$2.62$	&	$98.9$	&	$16.3$	&	$5.37$	&	$9.86$	&	$273$	&	$70$	&	$-52$	&	2.35 / 355	\\	
J0806	&	$5.89$	&	$93.0$	&	$4.6$	&	$6.58$	&	$2.63$	&	$420$	&	$55$	&	$-31$	&	1.00 / 194	\\	
RBS1223	&	$5.48$	&	$82.3$	&	$38.8$	&	$5.26$	&	$5.48$	&	$188$	&	$140$	&	$-151$	&	1.56 / 229	\\	
J1605	&	$6.33$	&	$102.4$	&	$6.3$	&	$6.27$	&	$10.52$	&	$437$	&	$44$	&	$-29$	&	1.88 / 282	\\	
J1856	&	$1.85$	&	$62.0$	&	$169.3$	&	$7.05$	&	$0.63$	&	--	&	--	&	--	&	1.19 / 206	\\	
RBS1774	&	$6.53$	&	$98.6$	&	$64.7$	&	$5.96$	&	$3.40$	&	$432$	&	$10$	&	$-12$	&	1.35 / 218	\\	\hline
 \end{tabular}}\label{tab:bbpl_par}
\begin{tabnote}
Norm. is the count rate at 1 keV in units of $10^{-5}$ s$^{-1}$ keV$^{-1}$, and $\Gamma$ is the photon index.
\end{tabnote}
\end{table}


\end{document}